\documentclass[pra, twocolumn, superscriptaddress, nofootinbib]{revtex4-2}
\usepackage{amssymb}
\usepackage{bbold}
\usepackage{hyperref}
\hypersetup{plainpages=false,colorlinks=true,linkcolor=blue, citecolor=blue, urlcolor=blue}

\makeatletter
\AtBeginDocument{\let\LS@rot\@undefined}
\makeatother 
\usepackage{graphicx}
\usepackage{amsmath}
\usepackage{color}
\usepackage{float}
\usepackage{placeins}
\usepackage{xcolor}
\pagecolor{white}
\newcommand\Sigmav{\mathbf{\Sigma}}
\newcommand\Gv{\mathbf{G}}
\newcommand\kv{\mathbf{k}}
\newcommand\rv{\mathbf{r}}

\begin{document}
\title{Topological superconductivity in a Hubbard model for twisted bilayer cuprates}
\author{T. Vibert}
\affiliation{D\'epartement de Physique, RQMP \& Institut Quantique, Universit\'e de Sherbrooke, Qu\'ebec, Canada J1K 2R1}
\author{D. S\'en\'echal}
\affiliation{D\'epartement de Physique, RQMP \& Institut Quantique, Universit\'e de Sherbrooke, Qu\'ebec, Canada J1K 2R1}

\begin{abstract}
We investigate the emergence of nontrivial topology in a twisted cuprate bilayer described by the Hubbard model in the weak-interaction regime. Our results show that the topological character depends sensitively on the doping level. For $U/t=3.85$, the Chern number assumes a value of $\pm 8$ in the electron-doped case, whereas it vanishes (0) in the hole-doped regime. The presence of nontrivial topology is further supported by an analysis of the associated edge states and their chirality in a finite-width geometry, while keeping full correlations.
\end{abstract}

\maketitle
\section{Introduction}

The possible observation of Majorana fermions has stimulated extensive research on possible topological superconductors~\cite{nadj2014observation, Machida2019, Zhang2018Science, Wang2022Review, Hosur2011, Hor2010, FuBerg2010}. The main interest in such systems lies in their potential use for realizing qubits that are robust against environmental perturbations~\cite{kitaev_unpaired_2001}. However, the interpretation of these edge states as Majorana modes remains controversial~\cite{dumitrescu_majorana_2015}, motivating further theoretical investigations and additional experimental studies. 

Given the paucity of intrinsic topological superconductors, an interesting proposal was put forth~\cite{can_high-temperature_2021} to construct a twisted bilayer of $d$-wave superconducting cuprates. A twist of nearly 45$^\circ$ might then induce, by proximity effect, a $d_{x^2-y^2}+id_{xy}$ superconducting state that breaks time-reversal invariance and has nontrivial topology. Given that superconductivity has been observed in monolayer Bi-2212 (Bi$_2$Sr$_2$CaCu$_2$O$_{8+\delta}$) \cite{yu_high-temperature_2019}, this proposal seems realizable in practice.

Theoretically, the analysis of Ref.~\cite{can_high-temperature_2021} has been performed at the mean-field level, within an effective model with attractive interactions.
However, cuprates are strongly-correlated materials and the superconducting pairing mechanism is based on the exchange of short-range spin fluctuations (see, e.g.,~\cite{kowalski2021}). The superconducting properties (critical temperature, order parameter, superfluid stiffness) are also strongly doping dependent.
Hence it is interesting, if not essential, to investigate the properties of twisted cuprate bilayers within a microscopic model that adequately describes high-temperature superconductivity in some universal fashion, even in the monolayer. The one-band Hubbard model is the simplest such model.
In Refs~\cite{lu_doping_2022, belanger_doping_2024,belanger2024a}, systems with $53.1^\circ$ and $43.6^\circ$ twists were studied in this fashion (these specific twist angles are necessary to produce commensurate systems amenable to the methods used in these works).
Even though these studies could identify doping ranges in which a time-reversal-breaking (TRB) superconducting solution exists, these solutions were not topological. This was attributed to the strong-coupling regime in which these systems were studied (beyond the Mott transition).
It is indeed known that strong-enough correlations can destroy topology~\cite{Imriska2016, Karnaukhov2026, Rachel2010, Amaricci2016, Wu2016}.

The present work aims to identify nontrivial topology in a simple Hubbard model for the twisted bilayer, the same as in Ref.~\cite{lu_doping_2022}, but this time in a weaker-interaction regime, below the Mott transition.
We find that there are indeed topological solutions in this regime, even though they are difficult to describe in traditional terms, because of the multiplicity of bands in the systems studied. We demonstrate the topological character of these solutions by computing the Chern number from the approximate Green function found via the variational cluster approximation (VCA), and also by identifying gapless edge modes in a finite-width system.

This paper is organized as follows: In Sec.~\ref{sec:model}, we review the model used; the superconducting pairing fields are adapted to the $D_{4}$ symmetry of the system, followed by a short review of the VCA, used to compute the superconducting order parameter. 
The computation of the Chern number is then discussed and generalized to the interacting case. 
In Sec.~\ref{sec:results}, we identify doping ranges in which a TRB superconducting solution exists and compute the Chern number from the Green function. Setting the interaction strength to $U=3.85$, just below the Mott transition, we find nontrivial topology with Chern number $\pm8$ in the electron-doped region, and trivial topology in the hole-doped region. This is confirmed by computing the spectral function in a finite-width (ribbon) geometry, in the fully correlated case.

\section{Model and method}
\label{sec:model}

Each layer of the bilayer system is described by the one-band Hubbard model. The complete Hamiltonian consists of one Hubbard Hamiltonian for each layer ($H^{(1,2)}$) and an interlayer Hamiltonian ($H_{\perp}$) containing the interlayer tunneling:
\begin{equation}
H = H^{(1)} + H^{(2)} + H_{\perp}.
\end{equation}
The Hubbard Hamiltonian for each layer is
\begin{align*}
H^{(\ell)} &= \kern-0.5em\sum_{\rv,\rv',\sigma}t_{\rv\rv'}c_{\rv\ell\sigma}^\dagger c_{\rv'\ell\sigma} 
+ U\sum_{\rv}n_{\rv\ell\uparrow}n_{\rv\ell\downarrow} 
- \mu\sum_{\rv,\ell,\sigma}n_{\rv\ell\sigma}~.
\end{align*}
Here, $c_{\rv\ell\sigma}$ is the electron annihilation operator for spin $\sigma$ at site $\rv$ of layer $\ell$, $n_{\rv\ell\sigma}$ the electron density operator, and $t_{\rv\rv'}$ the hopping amplitude, which include here first- ($t$) and second-neighbor ($t'$) hopping. $U$ is the on-site Coulomb repulsion and $\mu$ the chemical potential. We set $t=1$ as a unit of energy and $t'/t = -0.3$, a choice appropriate for 
Bi-2212.

\begin{figure}[!htp]
\begin{center}
\includegraphics[height=0.5\columnwidth]{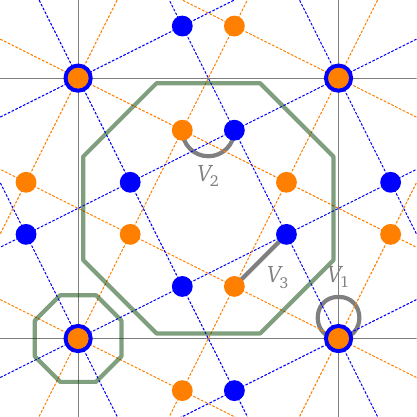}\hfil
\includegraphics[height=0.5\columnwidth]{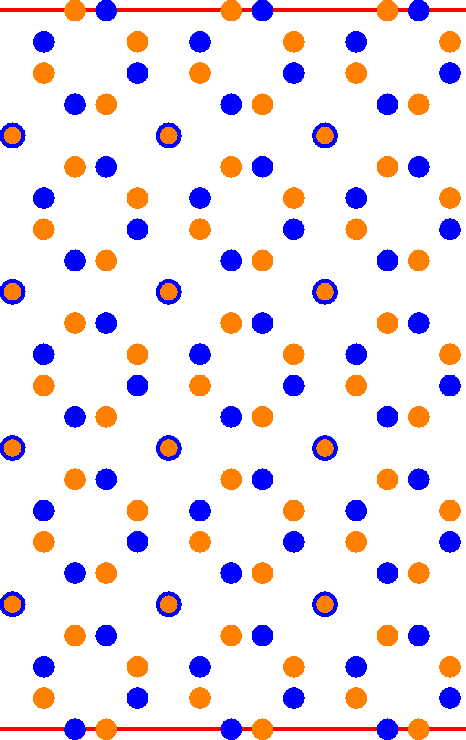}
\caption{Left panel: Unit cell of the twisted bilayer with a twist angle of $\theta = 2 \arctan\!\left(\frac{1}{2}\right)\sim 53.1^{\circ}$. Sites of the lower (upper) layer are shown in orange (blue). Interlayer tunneling terms are shown in gray. The two clusters used in the VCA, delimited by the green octagons, contain 8 and 2 sites, respectively.
Right panel: Ribbon configuration. The system is infinite along the $x$-axis and finite along the $y$-axis (here, the width is 5 unit cells only, but calculations were performed on a width of 40 unit cells). The red lines indicate the upper and lower edges.}
\label{fig:system}
\end{center}
\end{figure}

The interlayer Hamiltonian is
\begin{equation}
H_{\perp} = \sum_{n=1}^{3} V_{n} \sum_{\langle \rv,\rv' \rangle_{\perp,n},\sigma}
\left( c^{\dagger}_{\rv 1\sigma} c_{\rv' 2\sigma} + \mathrm{H.c.} \right)~~,
\end{equation}
where $V_{n}$ is the interlayer tunneling amplitude and $n = 1,2,3$ labels the type of interlayer coupling, as illustrated in Fig.~\ref{fig:system}. These amplitudes are defined relative to $V_{1}$ as $V_{2} = 0.5V_{1}$ and $V_{3} = 0.3V_{1}$.

This is of course an overly simplified description of the bilayer. 
In reality, each layer of Bi-2212 would be constituted of two copper-oxide planes.
Moreover, the interlayer tunneling amplitudes would likely be weaker and their precise values should be inferred \textit{ab initio}.
We are merely performing a proof of concept here, adopting parameter values that magnify the odds of finding topological solutions.

On each layer we define a $d$-wave superconducting pairing operator as follows:
\begin{align}
\label{eq:Weissd}
\hat{\Delta}^{(\ell)} &= \sum_{\rv\in\ell} \big[
c_{\rv\ell\uparrow}c_{\rv+\mathbf{x}\ell\downarrow}
- c_{\rv\ell\downarrow}c_{\rv+\mathbf{x}\ell\uparrow} \notag \\
&\quad - c_{\rv\ell\uparrow}c_{\rv+\mathbf{y}\ell\downarrow}
+ c_{\rv\ell\downarrow}c_{\rv+\mathbf{y}\ell\uparrow}
\big]~~.
\end{align}
Here, $\mathbf{x}$ and $\mathbf{y}$ denote the lattice vectors on each layer. 
However, these operators do not transform according to the representations of the $D_{4}$ symmetry of the bilayer (see Fig.~2 and Table~2 of Ref.~\cite{lu_doping_2022}).
To fix this, we introduce the following linear combinations:
\begin{equation}
\label{eq:WeissB1B2}
\hat{B}_{1} = \hat{\Delta}^{(1)} + \hat{\Delta}^{(2)} \qquad
\hat{B}_{2} = \hat{\Delta}^{(1)} - \hat{\Delta}^{(2)}~~.
\end{equation}
Each of these operators changes sign under a $\pi/2$ rotation with respect to the $z$ axis (perpendicular to the bilayer).
The $D_4$ group contains three different $\pi$-rotations about axes lying in the plane of the bilayer, and these two operators transform under the $B_1$ and $B_2$ representations of $D_4$.

The number of theoretical methods that can adequately treat the Hubbard model is quite limited. 
We use the variational cluster approximation (VCA)~\cite{Potthoff2003, dahnken2004, potthoff2014a}, a variational method based on Potthoff’s self-energy functional approach. 
The VCA builds upon cluster perturbation theory (CPT) \cite{senechal_spectral_2000} and can be regarded as its variational extension.
The first step is to tile the infinite lattice into identical clusters, and then to divide the Hamiltonian into a part $H'$ local to the clusters, plus an inter-cluster part $V$ : $H = H' + V$. The clusters must be small enough so that $H'$ can be solved numerically, via exact diagonalization, for the Green function $\Gv(\omega)$ and the self-energy $\Sigmav(\omega)$.
The clusters used in this work are the same as in Ref.~\cite{lu_doping_2022} and appear on Fig.~\ref{fig:system}.

The second step is to add to $H'$ (and consequently subtract from $V$) one-body operators that represent broken symmetry states.
In the current problem, these are the pairing operators \eqref{eq:WeissB1B2}, restricted to the clusters. Specifically, we shift
\begin{align}
H' \rightarrow H'
&+ \frac{1}{2}B_{1r} (\hat{B}_{1} + \hat{B}_{1}^{\dagger})
+ \frac{i}{2}B_{1i} (\hat{B}_{1} - \hat{B}_{1}^{\dagger}) \notag \\
&+ \frac{1}{2}B_{2r} (\hat{B}_{2} + \hat{B}_{2}^{\dagger})
+ \frac{i}{2}B_{2i} (\hat{B}_{2} - \hat{B}_{2}^{\dagger})~~.
\end{align}
where $B_1 = B_{1r}+iB_{1i}$ and $B_2 = B_{2r}+iB_{2i}$ are the ``Weiss fields'' that need to be set to some optimal value.
We stress that these are added to $H'$ only, not to $H$.

In order to set the Weiss fields to their optimal value, we use Potthoff's variational principle:
The Potthoff self-energy functional is defined as
\begin{equation}
\label{eq:Potthoffself}
\Omega_t (\Sigmav) = F(\Sigmav) - \text{Tr} \ln(-\Gv_{0t}^{-1} + \Sigmav)~~,
\end{equation}
where
\begin{equation}
\label{eq:F}
F(\Sigmav) = \Phi (\Gv) - \text{Tr}(\Sigmav \Gv)
\end{equation}
is the Legendre transform of $\Phi (\Gv)$, the Luttinger–Ward functional. Here, $\Sigmav$ is the self-energy, $\Gv$ the Green function, and $\Gv_{0t}$ the noninteracting Green function. 
The physical self-energy $\Sigma^*$ is obtained by applying the variational principle $\delta \Omega_{t} [\Sigmav] / \delta \Sigmav = 0$ and the value
$\Omega_{t}(\Sigmav^*)$ at this point is the physical grand potential $\Omega=E-\mu N$.

The key approximation in the VCA is to restrict the variational space to the physical self-energies of $H'$. 
The key point are that the interaction $U$ is the same in $H$ as in the \textit{reference Hamiltonian} $H'$ and that the functional $F(\Sigmav)$ is universal, i.e., is the same for $H$ and $H'$, which differ only via their non-interacting part, i.e., $\Gv_0$.
Hence the value of $F(\Sigmav)$ is known from the numerical solution of $H'$ and the expression for $\Omega’_t (\Sigmav)$ can be computed exactly in the form~\cite{dionne2023pyqcm}
\begin{align}
\label{eq:TruePotthoff}
\Omega_{t}(\Sigmav) &= \Omega'(\Sigmav) \notag \\
&\quad + \int \frac{d\omega}{2\pi i} \frac{L}{N}\sum_{\tilde{k}}
\mathop{\rm tr}\ln\big[\mathbf{1} - \mathbf{V}(\tilde{k})\Gv(\tilde{k},\omega)\big]~~.
\end{align}
Here, $\mathbf{V}(\tilde{k})$ is the intercluster hopping matrix, basically the difference between $\Gv_0^{-1}(H)$ and $\Gv_0^{-1}(H')$, and
$\Omega'(\Sigmav)$ is the (known) grand potential of $H'$. 
Since the self-energy depends on the Weiss fields $B_{1,2}$, the Potthoff functional can be treated directly as a function of these. 
A stationary point satisfying $\partial\Omega_{t}(B)/\partial B_{1,2} = 0$ at a nonzero Weiss field value signals the presence of superconductivity in the system.

The main purpose of using VCA is to determine the superconducting order parameters and identify regions where the orders of types $B_1$ and $B_2$ coexist.
To assess whether the system is topological, we need to compute a topological invariant, namely the Chern number.
In a noninteracting system, or in a mean-field description of the interacting system, the Chern number is obtained from the Berry curvature $F(\kv)$:
\begin{equation}\label{eq:Chern}
C = 2\pi \int \frac{d^2 k}{(2\pi)^2} \, F(\kv) \qquad
F(\kv) = \frac{\partial A_y}{\partial k_x} - \frac{\partial A_x}{\partial k_y}
\end{equation}
where the Berry connection is 
\begin{equation}\label{eq:Berry_connection}
A_{x,y} (\kv) = -i \sum_{\epsilon_\alpha(\kv)<0} \langle\alpha, \kv| \partial_{k_{x,y}} |\alpha,\kv\rangle~~.
\end{equation}
The sum is taken over the different occupied bands, indexed by $\alpha$.
In a simple two-band model with spin-orbit interaction, or triplet superconductivity, the kets $|\kv\rangle$ are the eigenvectors of a $\kv$-dependent, $2\times2$ Hermitian matrix and the Chern number is the index of the mappings from the torus (the Brillouin zone) to the unit sphere.
In the present case, the situation is more complex, since we are dealing with 20 different bands (twice the number of sites within the unit cell) and the Chern number does not have an obvious index interpretation in terms of simple manifolds, but rather as an index over a Grassmannian manifold $U(20)/(U(m)\times U(20-m))$~\cite{Ryu2010,Chiu2016Classification}.

Moreover, in the interacting case, the definition \eqref{eq:Chern} cannot be applied directly since it relies on a noninteracting Hamiltonian. Instead, we use a Green-function-based topological invariant:
\begin{equation}\label{eq:ChernG}
N = \frac{1}{24\pi^{2}} \int d^{3}k \,
\mathrm{Tr}\!\Big[
\epsilon^{\alpha\beta\gamma}
G \partial_{\alpha} G^{-1}
G \partial_{\beta} G^{-1}
G \partial_{\gamma} G^{-1}
\Big]~~,
\end{equation}
where $\alpha, \beta, \gamma \in \{0,1,2\}$ and 
$\partial_{0} = \partial / \partial \omega$, 
$\partial_{1} = \partial / \partial k_{x}$, 
$\partial_{2} = \partial / \partial k_{y}$. 
The connection between this invariant and the Chern number is discussed in~\cite{qi2008, Lovesey02102014, lessnich2023}. 
Integrating \eqref{eq:ChernG} over frequency yields an expression similar to \eqref{eq:Chern}, except that the non-interacting Hamiltonian is replaced by the so-called topological Hamiltonian~\cite{wang_strongly_2012}:
\begin{equation}\label{eq:topoH}
h_{t}(\kv) = - G^{-1}(\kv, \omega = 0).
\end{equation}
The invariant \eqref{eq:ChernG} is well-defined if $G$ has no zeros or poles at $\omega=0$; topological transitions hence occur when a Fermi surface appears ($G^{-1}\to 0$), or at a Mott transition ($G\to 0$).
In general, the Green function in Eq.~\eqref{eq:topoH} is not exactly equal to $\Gv_{0t}^{-1} - \Sigmav$; it is rather a ``declusterized'', or rather ``periodized'' version of it~\cite{dionne2023pyqcm}.
In this work this makes no difference, since clusters do not extend beyond one unit cell.
The integral \eqref{eq:Chern} over wave vectors is performed on a two-dimensional numerical grid covering the Brillouin zone. 
We use $10^{6}$ grid points, with an adaptive resolution: regions where the Berry curvature varies strongly are recursively subdivided, thereby improving accuracy without overly increasing the total number of grid points.

Finally, let us comment on the cluster geometry, illustrated on the left panel of Fig.~\ref{fig:system}.
The unit cell is divided into two clusters: one made of two superimposed sites and another of eight sites forming an octagon. This way each of the two clusters has the symmetry of the bilayer ($D_4$) and so has the approximate Potthoff functional \eqref{eq:TruePotthoff}. The Weiss fields $B_1$ and $B_2$ are only defined on this second cluster, but this is enough to propagate superconductivity to the whole system~\cite{lu_doping_2022}.

\section{Results}
\label{sec:results}

\begin{figure}[!htp]
\begin{center}
{\includegraphics[width=1\columnwidth]{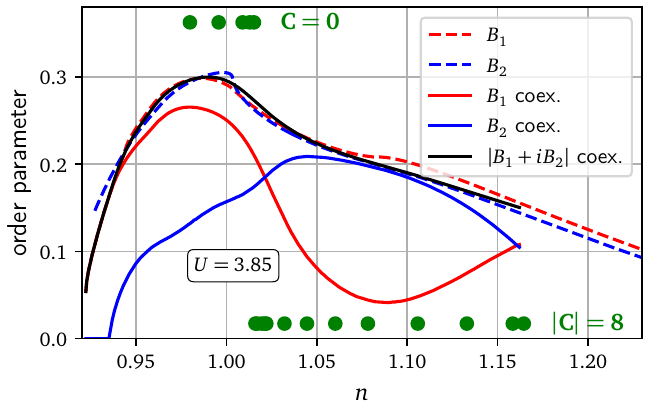}}
\caption{Order parameters obtained by VCA and Chern number $|C|$ (green dots) for \( U = 3.85 \) as a function of electronic density. Dashed lines represent order parameters calculated independently, while solid lines represent order parameters in the $B_1+iB_2$ phase. The Chern number was computed for a subset of the solutions.}
\label{fig:order_chern}
\end{center}
\end{figure}

In previous works on this system~\cite{lu_doping_2022, belanger_doping_2024}, the repulsive interaction was set to $U = 8$, in the strong-coupling regime.
The superconducting phase appears at finite doping and the order parameter goes to zero at half-filling (we ignore antiferromagnetism).
Suspecting that the Mott transition was the cause of the absence of topology, despite the presence of TRB superconductivity, we chose in this work a value of $U$ that lies below the Mott transition. At $U = 3.85$, the systems is superconducting at and around half-filling.
Figure~\ref{fig:order_chern} shows the order parameters in three cases: (i) superconductivity is only allowed to arise in the $B_1$ representation of $D_4$, (ii) the same in the $B_2$ representation and (iii) a homogeneous coexistence of both representations (mixed phase) is allowed, with a $\pi/2$ phase difference: $B_1+iB_2$. The latter breaks time-reversal symmetry and may have nontrivial topology.

In contrast to the strong-interaction regime, which we find for $U\geq4$, the order parameter does not vanish at half-filling, but is maximum there, and decreases with doping on both sides.
In the coexistence phase, the two amplitudes $B_1$ and $B_2$ are lower, but the combined amplitude $|B_1+iB_2|$ is comparable to the pure phase amplitudes. The $B_1$ component dominates towards hole-doping, whereas the $B_2$ component dominates towards electron doping.
Note that the distinction between hole and electron doping arises because of the second-neighbor hopping amplitude $t'$. If the sign of $t'$ were to change, the roles of electrons and holes would be interchanged.

The next step is to assess the topology of this TRB phase.
We compute the Chern number numerically according to the topological Hamiltonian \eqref{eq:topoH}, as explained at the end of the previous section.
The green dots on Fig.~\ref{fig:order_chern} show the computed Chern number, for a subset of solutions, as computed from the topological Hamiltonian~\eqref{eq:topoH}.
The Chern number vanishes below half-filling and takes the value $-8$ above half-filling, indicating that a nontrivial topological phase emerges only in the electron-doped regime (the sign is a matter of convention, and could equally be $+8$ if the complex order parameter $B_1+iB_2$ had the opposite but energetically equivalent chirality).

\begin{figure}[!htp]
\begin{center}
{\includegraphics[width=1\columnwidth]{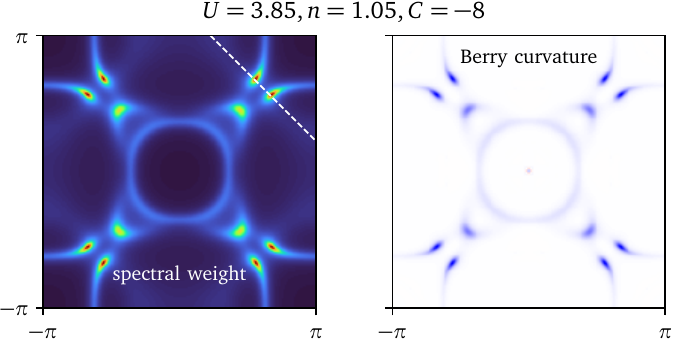}}
{\includegraphics[width=0.6\columnwidth]{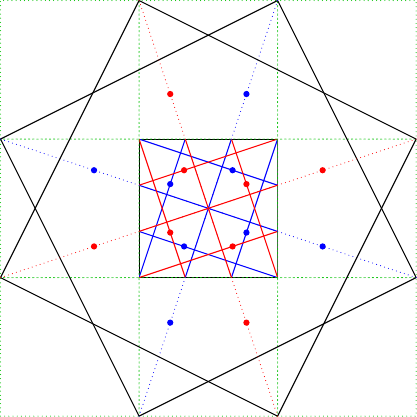}}
\caption{Top panel, left: Spectral weight of the bilayer at $\omega=0$ and $U=3.85$ in the electron-doped region (red is highest).
Top panel, right: Berry curvature $F$ (white=0), computed from the topological Hamiltonian and Eqs~(\ref{eq:Chern},\ref{eq:Berry_connection}); this integrates to $C=-8$ over the Brillouin zone.
Bottom panel: Folding of each layer's Brillouin zone into the bilayer Brillouin zone (center square), five times smaller.
The nodal lines are indicated in blue (lower layer) and red (upper layer).\label{fig:mdc_U385}}
\end{center}
\end{figure}

Computing the Chern number is a relatively delicate task from a numerical point of view because of the small size of the spectral gap, hence the small region of the Brillouin zone in which the vortex of the Berry connexion occurs.
To check the validity of this Chern number, we must make sure that the system is gapped. 
Figure~\ref{fig:mdc_U385} (top left panel) shows the spectral weight of the electron-doped system at the Fermi level ($\omega=0+i\eta$), for $U=3.85$ and $5\%$ doping.
Since a small but nonzero imaginary part $\eta=0.001$ is added, each pole in frequency ``spills'' somewhat over a small frequency interval and the resulting spectral function is nonzero even in a fully gapped state. The intensity in the plot simply represents where in the Brillouin zone one-particle states are close to the Fermi level.

We see from Fig.~\ref{fig:mdc_U385} that the low-energy states are localized at specific points in the Brillouin zone. 
The location of these points arises from the folding of the Brillouin zone due to the bilayer structure:
Each layer has its own Brillouin zone, represented by the black square in the bottom panel of Fig.~\ref{fig:mdc_U385}. 
The bilayer has a lower translational symmetry, hence a reduced Brillouin zone (the green square). 
The solid red and blue lines correspond to the $d$-wave nodal directions in each layer, folded into the reduced zone, while the dashed lines represent the same lines within the original zones. 
The red and blue points represent possible nodes on each layer, both in the original zones and folded in the reduced zone.
This folding scheme explains the presence of low-lying states located on each side of the diagonal on the top left panel of Fig.~\ref{fig:mdc_U385}. 
The additional low lying states along the diagonal arise because of correlation effects (they are absent from a mean-field state) and their position relative to the Fermi level depends on doping.

The top-right panel of Fig.~\ref{fig:mdc_U385} shows the Berry curvature, computed from Eqs~(\ref{eq:Chern},\ref{eq:Berry_connection}). It is highest where the low-lying states are concentrated, and is rather diffuse, meaning that it cannot be associated with precise singularities (e.g. nodes) in the Brillouin zone. It integrates to precisely $-8$ in this case.

\begin{figure}[!htp]
\begin{center}
{\includegraphics[width=1\columnwidth]{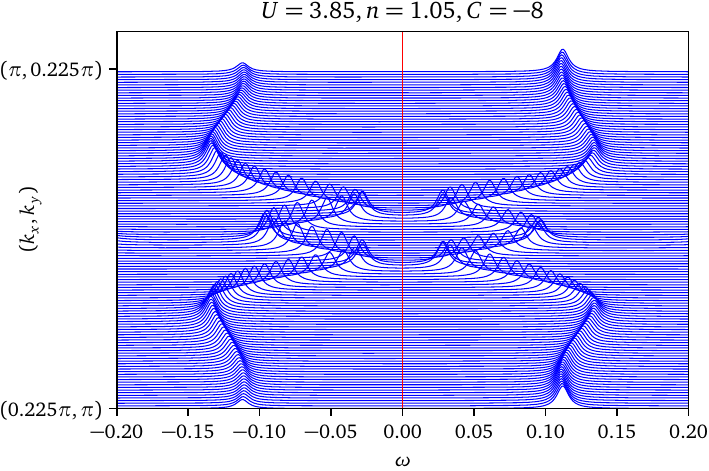}}
\caption{Spectral function along the wavevector path indicated on Fig.~\ref{fig:mdc_U385}, for the same system. This path is chosen so as to go through the nodal points and displays the smallest possible spectral gap.}
\label{fig:spectrum_2D_U385}
\end{center}
\end{figure}

To confirm the presence of a gap, we show on Fig.~\ref{fig:spectrum_2D_U385} the spectral function along a path chosen to maximize the spectral weight, indicated by a white dashed line on Fig.~\ref{fig:mdc_U385}. This reveals a gap even at the would-be nodal points and confirms that the Chern number calculated  at this value of the chemical potential $\mu$ is well-defined. This is also the case (not shown) at all densities where it was computed, for the green dots indicated on Fig.~\ref{fig:order_chern}.

Having established the existence of a gap, we next study the chirality of the edge states. For this purpose, the system is placed on a ribbon geometry, as shown in the right panel of Fig.~\ref{fig:system}. The spectral function is computed separately for the lower and upper edges, represented by the solid red lines in Fig.~\ref{fig:system}.
Each edge is constituted of two sites per unit cell (one per layer) and the edge spectral weight is the probability density that a particle, created or destroyed on each edge, have an energy $\omega$.

\begin{figure}[!htp]
\begin{center}
{\includegraphics[width=1\columnwidth]{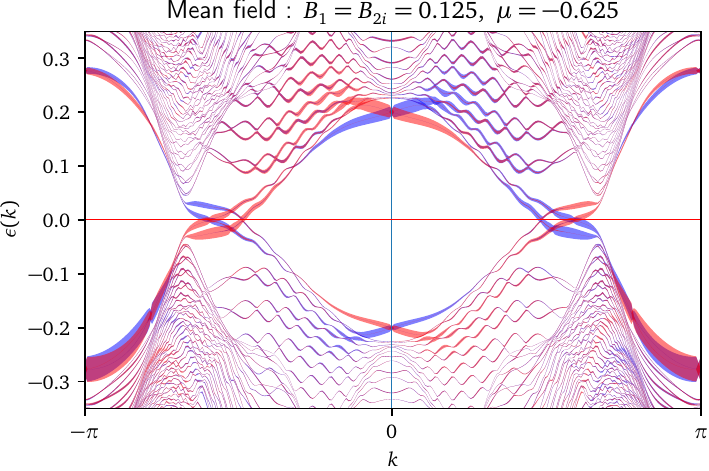}}
\caption{Edge-state dispersion of a ribbon of width $N=40$ unit cells in the mean-field approximation, with external pairing fields \( B_{1,2i} = 0.125 \). The thickness of each line reflects the weight of the corresponding edge sites contribution (blue for the lower edge, red for the upper edge). The number of crossings is compatible with a Chern number $C=4$.}
\label{fig:spectrum_edges_MFT}
\end{center}
\end{figure}

Let us start with the noninteracting case ($U = 0$), following the approach of Ref.~\cite{can_high-temperature_2021}.
We artificially set values to the pairing fields: $B_{1r}=B_{2i}=0.125$.
At $\mu= -0.625$, the Chern number computed from the bulk Green function is $C=4$. When the same system is studied in a ribbon geometry with $N$ unit cells in the $y$-direction, the $k_x$-dispersion of the $20N$ states are readily computed, and are shown in Fig.~\ref{fig:spectrum_edges_MFT} for $N=40$. 
The edge spectral weight is indicated by the thickness of each line (``fat bands''), in red (blue) for the upper (lower) edge. There are clearly four states crossing the Fermi level, with a clear chirality. Hence there is agreement between the boundary and bulk analyses. Note that even if the dispersions crossing the Fermi level correspond to proper edge states, the latter do not have a monopoly on the edge sites: Particles occupying bulk states also have a nonzero probability of sitting on the edge, as Fig.~\ref{fig:spectrum_edges_MFT} clearly shows.

\begin{figure}[!htp]
\begin{center}
\includegraphics[width=1\columnwidth]{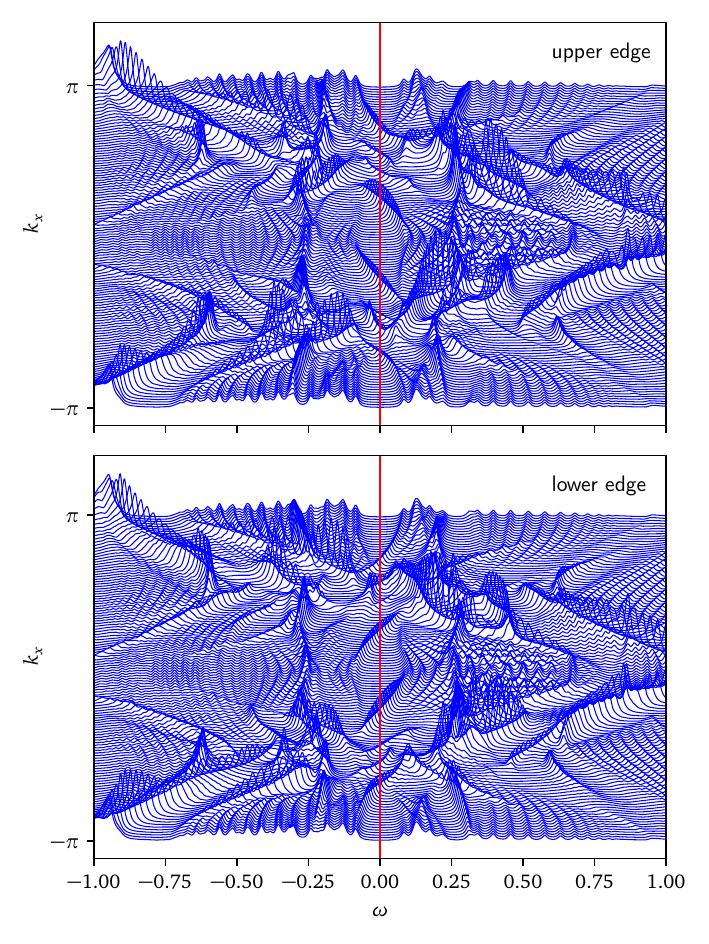}
\caption{Top panel : Spectral function on the upper edge sites in the ribbon geometry with a width of 40 unit cells at $U = 3.85$ and $\mu=1.58$, corresponding to a bulk electron doping of 5\% and a Chern number $|C|=8$. Bottom panel: the same, on the lower edge sites.}
\label{fig:spectrum_edges}
\end{center}
\end{figure}

Fig.~\ref{fig:spectrum_edges} now shows the edge spectral functions in the interacting case, with the same model parameters as previous figures.
A ribbon of width 40 unit cells was used for these spectral plots. Since the Chern number is $|C|=8$ in this case, as computed from the bulk Green function, we expect 8 dispersive peaks crossing the Fermi level and this is indeed what is observed (note that the energy window is wider than on Fig.~\ref{fig:spectrum_edges_MFT}).
However, the spectrum is that of a correlated system: the dispersing features cannot be uniquely ascribed to one-particle states, many secondary peaks are precursor to incoherent background, and so on. Nevertheless, the quasiparticle peaks crossing the Fermi level are evident, and so is chirality:
The spectral functions of the two edges are not symmetric under $k_x\to-k_x$, but are related by a rotation of $\pi$ in the $x$--$y$ plane, indicating that each edge supports distinct chiral edge states.

\begin{figure}[!htp]
\begin{center}
\includegraphics[width=1\columnwidth]{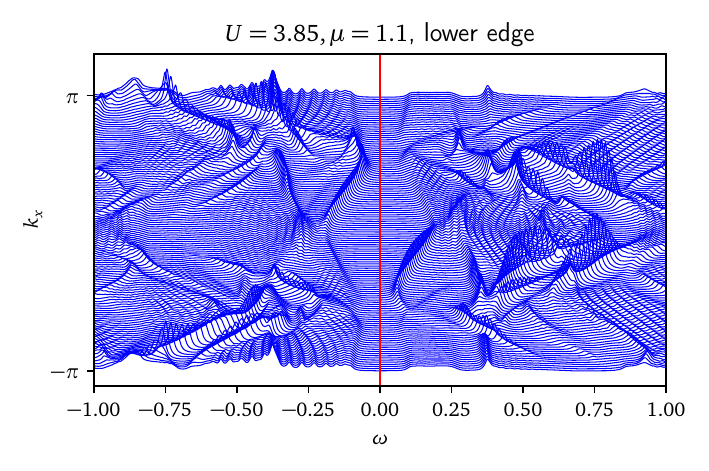}
\caption{Spectral function for the lower edge sites in the ribbon geometry with a width of 40 unit cells at $U = 3.85$ and $\mu=1.1$, corresponding to a  bulk hole doping of 2\% and a Chern number $C=0$. The upper edge is a mirror image when $k_x\to-k_x$. Note the presence of a gap.}
\label{fig:spectrum_edges_mu11}
\end{center}
\end{figure}

At last, Fig.~\ref{fig:spectrum_edges_mu11} shows the edge spectral function in the interacting case (for the lower edge only), with the same model parameters as in Fig.~\ref{fig:spectrum_edges}, except for the chemical potential ($\mu=1.1$) that now brings the mean occupation on the hole-doped side ($n\sim 0.98$), in the region of the TRB phase that has zero Chern number according to the bulk computation (Fig.~\ref{fig:order_chern}).
This time chiral edge states still exist (we only show the lower edge, the upper edge is identical except for a mirror symmetry $k_x\to-k_x$), but none of these states cross the Fermi level.
This confirms (i) the relation between edge states crossing the Fermi level and topology in the interacting case, and (ii) the lack of one-to-one relation between topology and the existence of a spontaneously broken time-reversal superconducting state.

How can we explain the doubling of the Chern number between the mean-field treatment ($C=4$, Fig.~\ref{fig:spectrum_edges_MFT}) and the correlated version ($|C|=8$, Fig.~\ref{fig:spectrum_edges})?
There is no simple way. 
The Berry curvature is derived from correlated bands; the structure of these low-energy bands is quite unlike that of the mean-field Hamiltonian: The topological Hamiltonian \eqref{eq:topoH} is obtained from the bare Hamiltonian by adding the self-energy at zero frequency, whereas the mean-field Hamiltonian would be derived by adding the self-energy at infinite frequency; there is no smooth connexion between the two.
Lowering the value of $U$ should make the mean-field approximation better, but on the other hand the VCA method used here would be less reliable. As $U$ is lowered, one can imagine that more correlated bands start contributing the the Chern number in compensating ways to lower its absolute value.

\section{Conclusion}

For a twisted cuprate bilayer described by the Hubbard model in the weak-interaction regime, we find a superconducting phase that breaks time-reversal symmetry in some range of densities using the variational cluster approximation.
In a doping range in the electron-doped region, the solutions are topological, as demonstrated by an explicit numerical computation of the Chern number derived from the effective topological Hamiltonian.
This is corroborated by the presence of a gap in the spectral function and by chiral edge states in a finite-width geometry, both consistent with the computed Chern number.
By contrast, other time-reversal symmetry breaking solutions have trivial topology (mostly on the hole-doped side).
This is a proof of principle that topological superconductivity can arise in this system when modeled microscopically, but only in the weak-coupling regime. This topology is destroyed in the strong-coupling regime, where the actual materials reside. This study shows that previous results obtained in the strong-coupling regime~\cite{lu_doping_2022}, where no topology arises, were not an artefact of the method used, but rather the effect of strong correlations.
A natural next step is to consider a more realistic model that captures the behavior of Bi-2212 across both weak- and strong-coupling regimes. Such a model should incorporate hopping parameters obtained \textit{ab initio} and explicitly include copper and oxygen orbitals.

\begin{acknowledgments}
Conversations with P. Fournier, E.~Lantagne-Hurtubise and A.-M. Tremblay are gratefully acknowledged. 
T.V. was supported by the Natural Sciences and Engineering Research Council (Canada) under Grant No. RGPIN-2020-05060.
\end{acknowledgments}

%

\end{document}